\documentclass{jpsj2}
\usepackage{graphicx}

\title{Metallic State of the Three-band Hubbard Model 
with Super-lattice Structure}

\author{Shigeru Koikegami$^1$
\footnote{E-mail: shigeru.koikegami@aist.go.jp}
, Takashi Yanagisawa$^1$, and Masaru Kato$^2$
}
\inst{$^1$Nanoelectronics Research Institute, 
AIST Tsukuba Central 2, Tsukuba 305-8568, Japan\\
$^2$Department of Mathematical Sciences, 
Osaka Prefecture University, Sakai 599-8531, Japan}

\recdate{January 27, 2006}

\abst{We investigate the dynamical superlattice correlation in the 
two-dimensional three-band Hubbard model on the basis of 
the unrestricted fluctuation exchange approximation. 
We calculate the one-particle spectral function, the spin correlation 
function and the charge correlation function at finite temperature. 
We find that some experimental results can be reproduced consistenly 
by taking inhomogenous distribution of Cu 3d electrons into account. 
The correlation functions suggest that several kinds of instabilities with 
spatial inhomogenieties exist in some regions, where these instabilities 
significantly affect the one-particle spectral fuctions.}

\kword{super-lattice structure, 
unrestricted fluctuation exchange approximation, 
strong correlation effect, spatial inhomogeneity}

\begin{document}
\sloppy
\maketitle

%
%
\section{Introduction}
The quasi-one-dimensional (Q1D) charge order, 
which is well-known as the stripe order, has been 
one of the significant issues in high-T$_{\mathrm{c}}$ cuprates (HTC) for 
recent years. The incommensurate peaks observed by the neutron 
scattering in La$_{\mathrm{2-x}}$Ba$_{\mathrm{x}}$CuO$_4$ 
are to be explained on the assumption that the static 
stripe order is present around 1/8-filling.\cite{Tranquada1995} The presence 
of the static stripe order in the ground state has been proved 
theoretically by the density matrix renormalization group method on the basis 
of the Q1D {\it t}-{\it J} model.\cite{White1998} 
Originally, since the beginning of HTC 
studies many experimentalists have pointed out the so-called 1/8-problem in 
La$_{\mathrm{2-x}}$Ba$_{\mathrm{x}}$CuO$_4$, i.e., the remarkable suppression 
of superconducting transition temperature at $x \sim 1/8$. 
Many theorists have 
thought that the strong on-site Coulomb interaction makes electrons 
almost localized with the long-period spin and charge correlations 
around 1/8-filling. They have studied these long-period correlations by the
 numerical or analytical methods on the basis of the 
two-dimensional (2D) Hubbard models and have clarified 
that some long-period correlated 
states can be the ground states around certain fillings.
\cite{Poilblanc1989,Zaanen1989,Kato1990,Salkola1996,Mizokawa1997,Machida1999,Ichioka1999,Kaneshita2001,Varlamov2002,Yanagisawa2002} These theoretically predicted long-period correlation 
states are consistent with 
the stripe state deduced from the neutron scattering experiments.\cite{Tranquada1999} 

The theoretical results on the basis of the 2D Hubbard models 
mean that the strong 
on-site interaction is enough to stabilize the stripe state near 1/8-filling 
as far as the ground state is considered. 
However, at finite temperature the strong 
fluctuations characteristic to the low dimensional strongly 
correlated electron systems exist. These strong fluctuations are composed of 
the many elements, e.g., the anti-ferromagnetic (AF) fluctuation and 
the charge ordering fluctuations with certain classes of super-lattice 
structures. These strong fluctuations 
accompany with the instabilities of the spin and charge orderings 
and simultaneously tend to interrupt them. 
The stripe with such strong fluctuations in metallic state 
is called a dynamical stripe.\cite{Zaanen1996,Zaanen2000,Zaanen2003,Hasselmann2002} 
The angle resolved photo-emission spectroscopy (ARPES) shows that 
the stripes in Bi$_2$Sr$_2$CaCu$_2$O$_{8+\delta}$ has more dynamic 
character than those in La$_{\mathrm{2-x}}$Sr$_{\mathrm{x}}$CuO$_4$. 
The discrimination whether the stripe is static or dynamic is to be 
judged from the characteristic energy of the fluctuation mode. 
If the fluctuation mode has low characteristic energy, 
i.e., the stripe fluctuates slowly, the mode 
can easily couple with phonon and the static stripe long-range 
order will appear. On the other hand, 
if the fluctuation mode has high characteristic energy, 
i.e., the stripe fluctuates fast, the mode cannot couple with phonon and 
the system will stay metallic together with the fluctuation. 
When we consider these fluctuations in the analysis of these models 
without the long-ranged Coulomb interaction, 
we can expect that the metallic state should be recovered 
at higher temperature. In this state, however, 
the electronic property is to be still affected by the stripe instability 
as far as its dynamical factor is investigated . 

In this paper, we analyze the 2D three-band Hubbard model 
on the basis of the unrestricted fluctuation exchange approximation (UFLEX). 
UFLEX can consider a strong correlation effect and a number of 
the spatial inhomogeneities without any assumptions about the ground state. 
These considered effects contain not only a static component 
but a dynamic one. As a result, the electronic state 
reproduced on the basis of UFLEX is affected by 
a number of spatially inhomogeneous fluctuations simultaneously. 
In fact, when we calculate the one-particle spectral functions 
and the spin correlation functions for our fully self-consistent solutions, 
we can find that the one-particle spectra of the electronic state 
reflect these exotic state accompanied with the anti-ferromagnetic 
or the further long-periodic instabilities. Especially, near $1/8$-filling 
the spin and charge correlation functions show the existence of the 
Q1D fluctuation at finite frequency. This fast Q1D fluctuation is originated 
from the striped spatial inhomogeneity. Thus, 
our results suggest that the spatial inhomogeneities should be concerned 
in the strongly correlated electron system when the electronic behavior 
at a finite frequency is discussed. 

%
%
\section{2D three-band Hubbard model}

Our 2D three-band Hubbard model Hamiltonian, $H$, is composed of {\it d}-electrons at each Cu site and 
{\it p}-electrons at O site. We consider only the on-site Coulomb repulsion $U$ between 
{\it d}-electrons at each Cu site. Then, $H$ is divided 
into the non-interacting part, $H_0$, and the interacting part, $H_1$, as 
\begin{equation}
H=H_0+H_1%
-\mu \sum_{{\mathbf k} \sigma}
(d_{{\mathbf k} \sigma}^{\dagger}d_{{\mathbf k} \sigma}
+p_{{\mathbf k} \sigma}^{x \dagger}p_{{\mathbf k} \sigma}^x
+p_{{\mathbf k} \sigma}^{y \dagger}p_{{\mathbf k} \sigma}^y).
\end{equation} 
Here $d_{{\mathbf k} \sigma}(d_{{\mathbf k} \sigma}^{\dagger})$ and 
$p_{{\mathbf k} \sigma}^{x(y)}(p_{{\mathbf k} \sigma}^{x(y) \dagger})$ are 
the annihilation (creation) operator for {\it p}- and 
{\it p}$^{x(y)}$-electron of momentum ${\mathbf k}$ and spin $\sigma$, 
respectively. $\mu$ is the chemical potential.
The non-interacting part $H_0$ is represented by 
\begin{equation}
H_0= {\sum_{{\mathbf k} \sigma}}
\left(
d_{{\mathbf k} \sigma}^\dagger \, 
p_{{\mathbf k} \sigma}^{x \dagger} \, 
p_{{\mathbf k} \sigma}^{y \dagger}
\right) \left(
\begin{array}{ccc} \Delta_{dp} & \zeta_{\mathbf k}^x  & \zeta_{\mathbf k}^y \\ 
-\zeta_{\mathbf k}^x & 0 & \zeta_{\mathbf k}^p \\
-\zeta_{\mathbf k}^y & \zeta_{\mathbf k}^p & 0 \\
\end{array} \right)\!
\left( \begin{array}{c} d_{{\mathbf k} \sigma} \\ p_{{\mathbf k} \sigma}^x
\\ p_{{\mathbf k} \sigma}^y \\
\end{array} \right),
\end{equation}
where $\Delta_{dp}$ is the hybridization gap energy
between {\it d}- and {\it p}-orbitals. 
We take the lattice constant of the square
lattice formed of Cu sites as the unit of length, and we can represent
$\zeta_{\mathbf k}^{x(y)}=2{\rm i}\,t_{dp} \sin \frac{k_{x(y)}}{2}$ and
$\zeta_{\mathbf k}^p=-4t_{pp} \sin \frac{k_x}{2} \sin \frac{k_y}{2}$,
where $t_{dp}$ is the transfer energy between a {\it d}-orbital and a
neighboring {\it p}$^{x(y)}$-orbital and $t_{pp}$ is that between a
{\it p}$^x$-orbital and a {\it p}$^y$-orbital. In this
study, we take $t_{dp}$ as the unit of energy. The residual part,
$H_1$, is described as
\begin{equation}
H_1= \frac{U}{N} \sum_{{\mathbf k} {\mathbf k}^\prime} \sum_{\mathbf q}
        d_{{\mathbf k}+{\mathbf q} \uparrow}^{\dagger} d_{{\mathbf
        k}^\prime- {\mathbf q} \downarrow}^{\dagger} d_{{\mathbf k}^\prime
        \downarrow} d_{{\mathbf k} \uparrow}.
\end{equation}
where $U$ is the on-site Coulomb repulsion between {\it d}-orbitals and
$N$ is the number of ${\mathbf k}$-space lattice points in the first
Brillouin zone (FBZ). 

%
%
\section{UFLEX}

We have developed UFLEX in order to analyze the spatially inhomogeneous 
system. In our UFLEX formulation we introduce 'cluster' momentum 
besides 'usual' momentum. The notion of this cluster momentum is 
suggested from the one in dynamical cluster approximation, 
in which the nonlocal correlations 
are considered beyond dynamical mean field theory.
\cite{Hettler1998,Maier2000,Imai2002}

However, our cluster momentum is 
not directly related with a coarse graining of the Brillouin zone. When 
the most stable state of our system is spatially homogeneous, only a cluster 
momentum, ${\mathbf K}={\mathbf 0}$, is enough to obtain 
the {\it true} solution for our problem. In that case, 
UFLEX is consistent to 'usual' FLEX.\cite{Bickers1989} 
The advantage of our introduction of cluster momenta is 
its flexibility in the correspondence to the problem. 
If the problem have some important classes of the spatial inhomogeneities, 
we will only need to provide the cluster momenta 
corresponding to those classes. For example, when 
the most important instability is the anti-ferromagnetic one, 
two cluster momenta, ${\mathbf K}={\mathbf 0}$ and ${\mathbf K}
=(\frac{\pi}{2},\frac{\pi}{2})$, are enough. Of course, the stripe 
instabilities have their characteristic 
cluster momenta. Thus, when we anticipate to encounter 
the stripe instabilities in our problem, we need to provide 
only those cluster momenta.

\begin{figure}
\includegraphics[width=6.1cm]{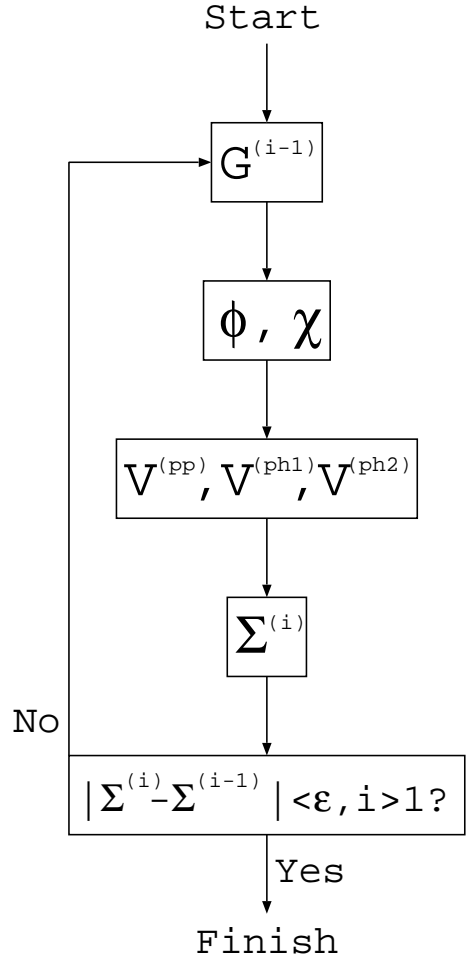}
\caption{The self-consistent procedure in UFLEX. 
$i$ represents the iteration count, and $\varepsilon$ is the small value.}
\label{figure:10}
\end{figure}

First, we define the unrestricted perturbed Green function, 
$G_{d\,\mathbf{K}}^{\,\sigma}(\mathbf{k},\mathrm{i}\epsilon_n)$.
Here we use the abbreviation of Fermion Matsubara frequencies, 
$\epsilon_n=\pi T(2n+1)$ with integer $n$, where $T$ is the temperature. $\mathbf{K}$ indicates 
the cluster momentum. The set, $\{\mathbf{K}\}$, is chosen so that 
any sum, $\mathbf{k}+\mathbf{K}$, exists in the set, $\{\mathbf{k}\}$. 
$G_{d\,\mathbf{K}}^{\,\sigma}(\mathbf{k},\mathrm{i}\epsilon_n)$ is calculated by the Dyson equation :
\begin{equation}
\left[G_{d\,\mathbf{K}}^{\,\sigma}(\mathbf{k},\mathrm{i}\epsilon_n)
\right]^{-1}=\delta_{\mathbf{K}} \cdot 
\left\{G_d^{\,\sigma\,(0)}
(\mathbf{k},\mathrm{i}\epsilon_n) \right\}^{-1}
-\Sigma_{\mathbf{K}}^\sigma(\mathbf{k},\mathrm{i}\epsilon_n), \\
\label{eq:1}
\end{equation}
where $\delta_{\mathbf{K}}$ means Kronecker's delta, 
i.e. $\delta_{\mathbf{K}=\mathbf{0}}=1$ 
and $\delta_{\mathbf{K} \neq \mathbf{0}}=1$.
$\left[\cdots\right]^{-1}$ indicates 
the inverse operation defined so as to satisfy the identity : 
\begin{equation}
\delta_{\mathbf{K}}= \sum_{\mathbf{K}^\prime}G_{d\,\mathbf{K}-\mathbf{K}^\prime}^{\,\sigma}
(\mathbf{k}+\mathbf{K}^\prime-\mathbf{K},\mathrm{i}\epsilon_n)
\left[G_{d\,\mathbf{K}^\prime}^{\,\sigma}(\mathbf{k},\mathrm{i}\epsilon_n)\right]^{-1}
\end{equation}
for all ${\mathbf{k}}$ and $n$.  $G_d^{\,\sigma\,(0)}(\mathbf{k},\mathrm{i}\epsilon_n)$ 
in Eq.~(\ref{eq:1}) is the unperturbed Green function derived by diagonalizing of $H-H_1$ as 
\begin{align}
& \left\{G_d^{\,\sigma\,(0)}(\mathbf{k},z-\mu)\right\}^{-1}= \nonumber \\
& z-\Delta_{dp}-\frac{2z(2-\cos k_x-\cos k_y)+
      8t_{pp}(1-\cos k_x)(1-\cos k_y)}
{z^2-4t_{pp}^2(1-\cos k_x)(1-\cos k_y)}.
\end{align}
In order to estimate our unrestricted self-energy, $\Sigma_{\mathbf{K}}^\sigma
(\mathbf{k},\mathrm{i}\epsilon_n)$, in Eq.~(\ref{eq:1}), 
we adopt the UFLEX as follows.
\begin{align}
& \Sigma_{\mathbf{K}}^\sigma
(\mathbf{k},\mathrm{i}\epsilon_n) = \nonumber \\
& \frac{TU}{N}\sum_{\mathbf{q}\,n^\prime}
G_{d\,\mathbf{K}}^{\,-\sigma}(\mathbf{k}-\mathbf{q},
\mathrm{i}\epsilon_{n^\prime})\left.e^{ \mathrm{i}\epsilon_{n^\prime}\eta}
\right|_{\eta \rightarrow +0} \nonumber \\
& +\frac{T}{N}
\sum_{\mathbf{q}\,\mathbf{K}^\prime\,m}
G_{d\,\mathbf{K}-\mathbf{K}^\prime}^{\,\sigma}
(\mathbf{q}-\mathbf{k}+\mathbf{K}^\prime-\mathbf{K},
\mathrm{i}\omega_m-\mathrm{i}\epsilon_n)
V_{\mathbf{K}^\prime}^{(\mathrm{pp})}
(\mathbf{q},\mathrm{i}\omega_m)
\nonumber \\
& +\frac{T}{N}
\sum_{\mathbf{q}\,\mathbf{K}^\prime\,m}\left[
G_{d\,\mathbf{K}-\mathbf{K}^\prime}^{\,-\sigma}(\mathbf{k}-\mathbf{q},
\mathrm{i}\epsilon_n-\mathrm{i}\omega_m)
V_{\mathbf{K}^\prime}^{\sigma\,(\mathrm{ph1})}(\mathbf{q},\mathrm{i}\omega_m)
\right. \nonumber \\
& \hspace{5.5em}+\left. G_{d\,\mathbf{K}-\mathbf{K}^\prime}^{\,\sigma}(\mathbf{k}-\mathbf{q},
\mathrm{i}\epsilon_n-\mathrm{i}\omega_m)
V_{\mathbf{K}^\prime}^{\sigma\,(\mathrm{ph2})}(\mathbf{q},\mathrm{i}\omega_m)
\right],
\label{eq:2}
\end{align}
\begin{align}
\label{eq:4}
& V_{\mathbf{K}}^{(\mathrm{pp})}(\mathbf{q},\mathrm{i}\omega_m) = \nonumber \\ 
& U^2\phi_{\mathbf{K}}(\mathbf{q},\mathrm{i}\omega_m) \nonumber \\ 
& -U^2\sum_{\mathbf{K}^\prime}
\phi_{\mathbf{K}-\mathbf{K}^\prime}(\mathbf{q},\mathrm{i}\omega_m)
\left[\,
\delta_{\mathbf{K}^\prime}
+U\phi_{\mathbf{K}^\prime}
(\mathbf{q}+\mathbf{K}-\mathbf{K}^\prime,\mathrm{i}\omega_m)\right]^{-1}, \\
\label{eq:5}
& V_{\mathbf{K}}^{\sigma(\mathrm{ph1})}(\mathbf{q},\mathrm{i}\omega_m) = \nonumber \\
& -U^2\chi_{\mathbf{K}}^{\sigma\, -\sigma}
(\mathbf{q},\mathrm{i}\omega_m) \nonumber \\ 
&+U^2\sum_{\mathbf{K}^\prime}
\chi_{\mathbf{K}-\mathbf{K}^\prime}^{\sigma\, -\sigma}
(\mathbf{q},\mathrm{i}\omega_m)
\left[\delta_{\mathbf{K}^\prime}
-U\chi_{\mathbf{K}^\prime}^{\sigma\, -\sigma}
(\mathbf{q}+\mathbf{K}-\mathbf{K}^\prime,\mathrm{i}\omega_m)
\right]^{-1}, 
\end{align}
and
\begin{align}
\label{eq:6}
& V_{\mathbf{K}}^{\sigma \,(\mathrm{ph2})}(\mathbf{q},\mathrm{i}\omega_m) = \nonumber \\
& U^2\sum_{\mathbf{K}^\prime}
\chi_{\mathbf{K}-\mathbf{K}^\prime}^{-\sigma -\sigma}
(\mathbf{q},\mathrm{i}\omega_m) \nonumber \\
& \times \left[\,\delta_{\mathbf{K}^\prime}\right.
-\!U^2\!\sum_{\mathbf{K}^{\prime \prime}}
\chi_{\mathbf{K}^\prime-\mathbf{K}^{\prime \prime}}^{\sigma \sigma}
(\mathbf{q}+\mathbf{K}-\mathbf{K}^\prime,\mathrm{i}\omega_m) \nonumber \\
& \hspace{7em}
\times \left. \chi_{\mathbf{K}^{\prime \prime}}^{-\sigma -\sigma}
(\mathbf{q}+\mathbf{K}-\mathbf{K}^{\prime \prime},\mathrm{i}\omega_m)
\right]^{-1},
\end{align}
where $\omega_m=2m\,\pi T$ with integer $m$ are Boson Matsubara frequencies, 
\begin{align}
\label{eq:3}
& \phi_{\mathbf{K}}(\mathbf{q},\mathrm{i}\omega_m)= \nonumber \\ 
& \frac{T}{N}\,\sum_{\mathbf{k} \mathbf{K}^\prime\, n}
G_{d\,\mathbf{K}-\mathbf{K}^\prime}^{\,\sigma}(\mathbf{q}-\mathbf{k}, 
\mathrm{i}\omega_m-\mathrm{i}\epsilon_n)
\,G_{d\,\mathbf{K}^\prime}^{-\sigma}(\mathbf{k},\mathrm{i}\epsilon_n),
\end{align}
and
\begin{align}
\label{eq:7}
& \chi_{\mathbf{K}}^{\sigma \sigma^\prime}(\mathbf{q},\mathrm{i}\omega_m)= \nonumber \\ 
& -\frac{T}{N}\sum_{\mathbf{k} \mathbf{K}^\prime\,n}
G_{d\,\mathbf{K}-\mathbf{K}^\prime}^{\sigma}
(\mathbf{q}+\mathbf{k},\mathrm{i}\omega_m+\mathrm{i}\epsilon_n)
\,G_{d\,\mathbf{K}^\prime}^{\sigma^\prime}(\mathbf{k}-\mathbf{K}^\prime
,\mathrm{i}\epsilon_n) .
\end{align}
In order to obtain the solution satisfying the conserving law, 
we need to solve 
Eqs.~(\ref{eq:1},\ref{eq:2},\ref{eq:4},\ref{eq:5},\ref{eq:6},\ref{eq:3},
\ref{eq:7}) fully self-consistently. This self-consistent procedure is 
shown in Fig.~\ref{figure:10} diagrammatically.

In numerical calculations we divide the FBZ into $16 \times 16$ meshes. 
We take $8\times 8$ cluster momenta so that we can reproduce a 
homogeneous state, inhomogeneous states 
with 2-, 4-, or 8-lattice period charge density wave (CDW) 
or spin density wave (SDW) along 
$a(x)$-axis or $b(y)$-axis, or an AF state. We prepare $2^{11}=2048$ 
Matsubara frequencies for temperature $T=0.020 \sim 180{\mathrm{K}}$. 
The other parameters :
$t_{dp}=1.0 \sim 0.80\,{\mathrm{eV}}$, 
$t_{pp}=0.60 \sim 0.48\,{\mathrm{eV}}$, and 
$\Delta_{dp}=0.70 \sim 0.56\,{\mathrm{eV}}$, 
$U=4.0 \sim 3.2\,{\mathrm{eV}}$, which are all common for our results.

%
%
\section{Fermi surface evolution with hole- or electron-doping}
\label{Fermi}

We have obtained the fully self-consistent solutions for 
$0.000 \leq \delta_{\mathrm{h}} \leq 0.057$, 
$0.112 \leq \delta_{\mathrm{h}} \leq 0.231$, 
and $0.000 \leq \delta_{\mathrm{e}} \leq 0.249$, 
where $\delta_{\mathrm{h}} \equiv 1-n_{\mathrm{d}}-n_{\mathrm{p}}$ and   
$\delta_{\mathrm{e}} \equiv n_{\mathrm{d}}+n_{\mathrm{p}}-1$. 
$n_{\mathrm{d}}$ and $n_{\mathrm{p}}$ represent the number of 
$d$- and $p$-electrons, respectively. 
We could not obtain any convergent solution for 
$0.057 < \delta_{\mathrm{h}} < 0.112$. According to the elaborated neutron scattering experiments on La$_{\mathrm{2-x-y}}$Ba$_{\mathrm{x}}$Sr$_{\mathrm{y}}$CuO$_4$,\cite{Fujita2002PRB,Fujita2002PRL,Tranquada2004} 
in this region the inhomogeneous states with longer 
than 8-lattice period SDW along diagonal, which is called diagonal 
stripe state, could be realized. Our present 
calculation would be short of provided cluster momenta to reproduce 
the inhomogeneous states with the diagonal fluctuations. Hereafter, we 
restrict our discussion to the states with the vertical fluctuations. 

In this section we investigate the Fermi 
surface evolution with doping in detail. In order to determine the Fermi 
surface, we define the one-particle spectral 
weight as 
\begin{equation}
A^{\sigma}(\mathbf{k},E) \equiv -\frac{1}{\pi}\mathrm{Im}
\sum_{\zeta=d,\,p^x,p^y}\left.
G_{\zeta\,\mathbf{0}}^\sigma(\mathbf{k},\mathrm{i}\epsilon_n)
\right|_{\mathrm{i}\epsilon_n \rightarrow E},
\label{eq:8}
\end{equation}
where
\begin{align}
& G_{p^{x,y}\,\mathbf{0}}^\sigma(\mathbf{k},z-\mu) \nonumber \\
&  = \left\{z[z-\Delta_{dp}-\Sigma_{\mathbf{0}}^\sigma(\mathbf{k},z)]-2(1-\cos k_{x,y})\right\} \nonumber \\ 
& \hspace{1em}\times\!\left\{[z^2-4t_{pp}^2(1-\cos k_x)(1-\cos k_y)][z-\Delta_{dp}-\Sigma_{\mathbf{0}}^\sigma(\mathbf{k},z)]\right. \nonumber \\
& \hspace{2.5em} 
\left.-2z(2-\cos k_x-\cos k_y)\!-\!8t_{pp}(1-\cos k_x)(1-\cos k_y)\right\}
^{\!-1}. 
\label{eq:9}
\end{align}
In Eq.~(\ref{eq:8}) we use Pad$\acute{\mathrm{e}}$ approximation 
for the method of analytic continuation. In order to 
determine the Fermi surface on the basis of our calculated results, 
we should calculate the one particle spectral weight at $E=0$, 
$A^{\sigma}(\mathbf{k},0)$, and find the $\mathbf{k}$-points at which 
$A^{\sigma}(\mathbf{k},0)$ becomes large. 
Because the thermal fluctuation with CDW, SDW, or anti-ferromagnetic 
instabilities can generate the branched energy dispersion, we show 
our calculated results as follows. We choose the both $\mathbf{k}$-points 
on which the one-particle spectral weight at $E=0$ has 
the largest and the second largest peaks for fixed $k_x$ 
as far as $0 < k_y \leq k_x < \pi$. And we repeat this operation for the other 
parts of FBZ. In the following figures, the points on which 
the one-particle spectral weight at $E=0$ has
the largest and the second largest peaks are to be 
indicated with black and gray circles, respectively. 
Our results have been obtained at high temperature, and the 
thermal fluctuations make the Fermi level vague. 
However, if we have empty spaces surrounded by these points, 
we can expect the branched energy dispersion near the Fermi level. The 
branched energy dispersion is made by a spatial inhomogeneous instability. 
Furthermore, these points can suggest the Fermi surface connectivities at lower 
temperature. 

\begin{figure}
\includegraphics[width=8.6cm]{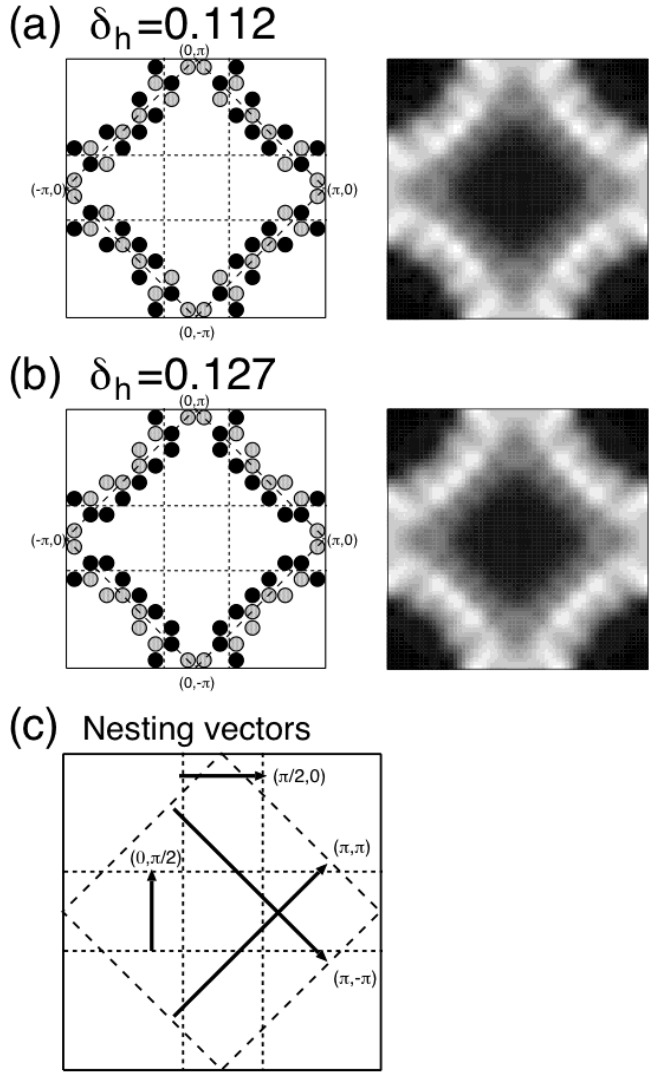}
\caption{Fermi surface evolutions near $1/8$-filling. 
Left sides : (a) for $\delta_{\mathrm{h}}=0.112$. 
(b) for $\delta_{\mathrm{h}}=0.127$. 
Right sides : The gray-scale images interpolated from raw data are attached 
on account of the guide to the eye. The brighter spots have the larger spectral weights. (c) Illustration of nesting vectors. Dashed lines represent the planes connected with one another by the nesting vectors, $(\pi,\pi)$ or $(\pi,-\pi)$. Dotted lines represent the planes connected with one another by the nesting vectors, $(\pi/2,0)$ or $(0,\pi/2)$. 
The illustration of nesting vectors 
is also applicable to the other figures representing 
Fermi surface evolutions.}
\label{figure:1}
\end {figure}
Firstly, we should mention our results for the hole-doped states 
around $1/8$-filling, in which the dynamical stripe state might be realized. 
As shown in Fig.~\ref{figure:1}, 
near $1/8$-filling we have eight empty spots around the intersections 
of the planes indicated by dashed and dotted lines, which relate 
to the anti-ferromagnetic and the 4-lattice period 
CDW instabilities, respectively. 
It means that the one-particle spectral weights are lost due 
to the fluctuations both of the anti-ferromagnetic instability and of 
the 4-lattice period CDW instability. 
These instabilities occur when some parts of the Fermi surface are 
connected with one another by the nesting vectors. The parts of Fermi 
surface connected with one another by the nesting vectors, $(\pi,\pi)$ 
or $(\pi,-\pi)$, are called 'hot spots'. On hot spots the quasi-particles 
play important roles in the transport phenomena. 
Therefore, we expect some anomalous electronic behaviors near $1/8$-filling, 
when the one-particle spectral weights are lost on hot spots. 
This is the reason why so-called $1/8$-problem occurs around $1/8$-filling. 

\begin{figure}
\includegraphics[width=8.6cm]{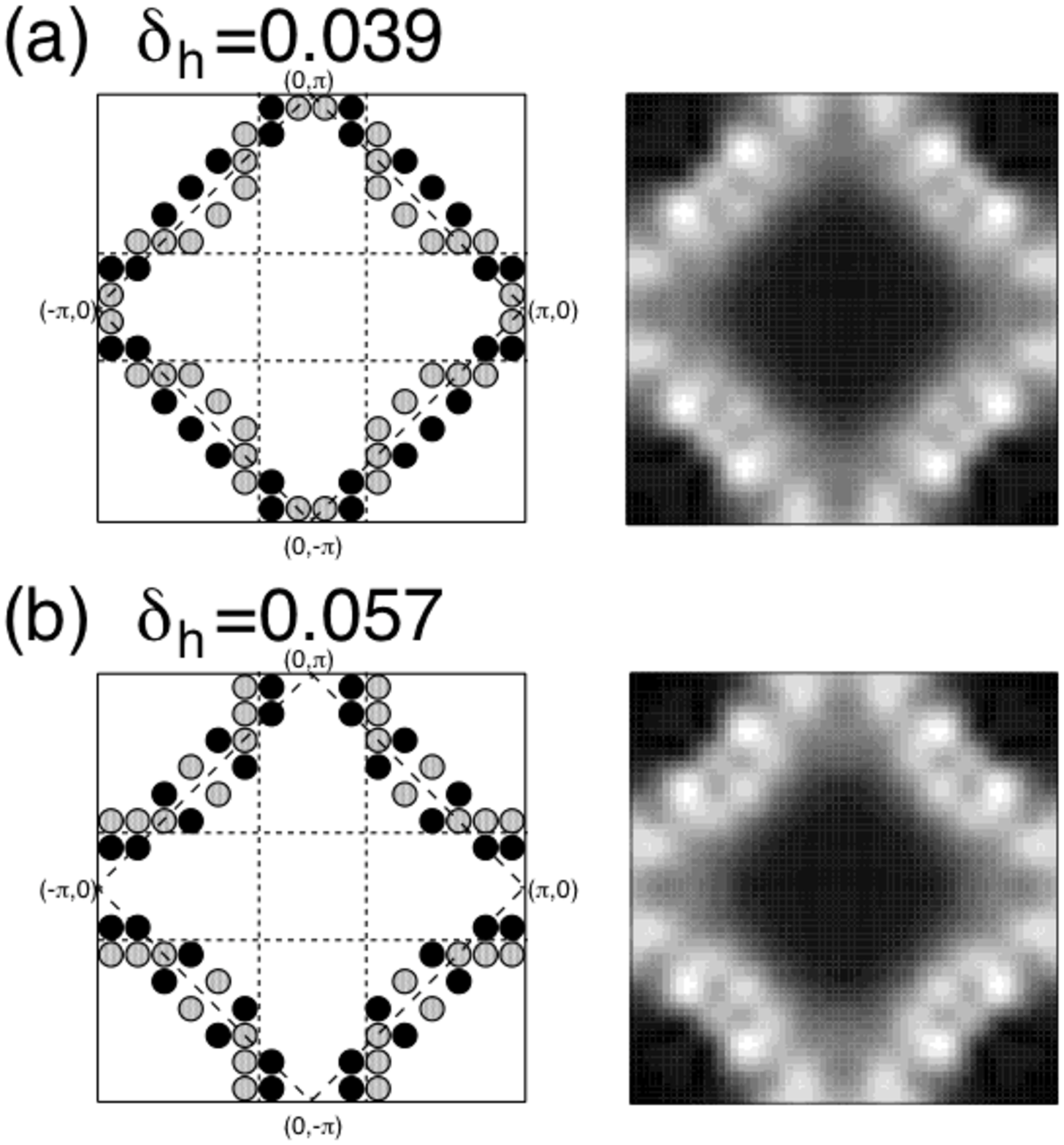}
\caption{Fermi surface evolutions for lightly-hole-doped states. (a) for $\delta_{\mathrm{h}}=0.039$. (b) for $\delta_{\mathrm{h}}=0.057$.}
\label{figure:2}
\end{figure}
\begin{figure}
\includegraphics[width=8.6cm]{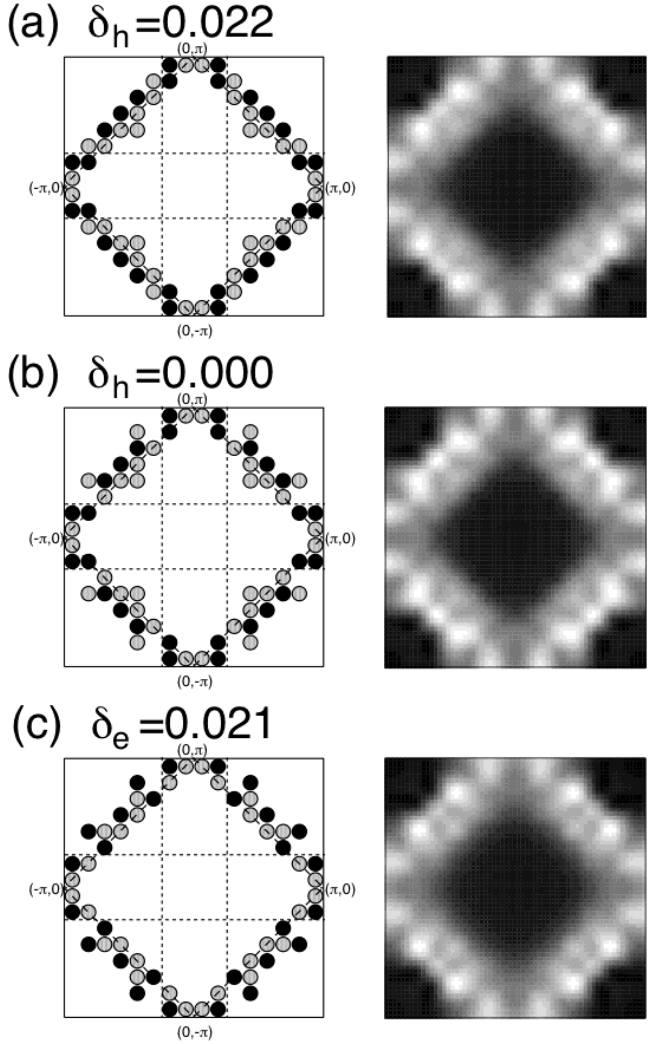}
\caption{Fermi surface evolutions for near half-filling. (a) for $\delta_{\mathrm{h}}=0.022$. (b) for half-filling. (b) for $\delta_{\mathrm{e}}=0.021$.}
\label{figure:4}
\end {figure}
\begin{figure}
\includegraphics[width=8.6cm]{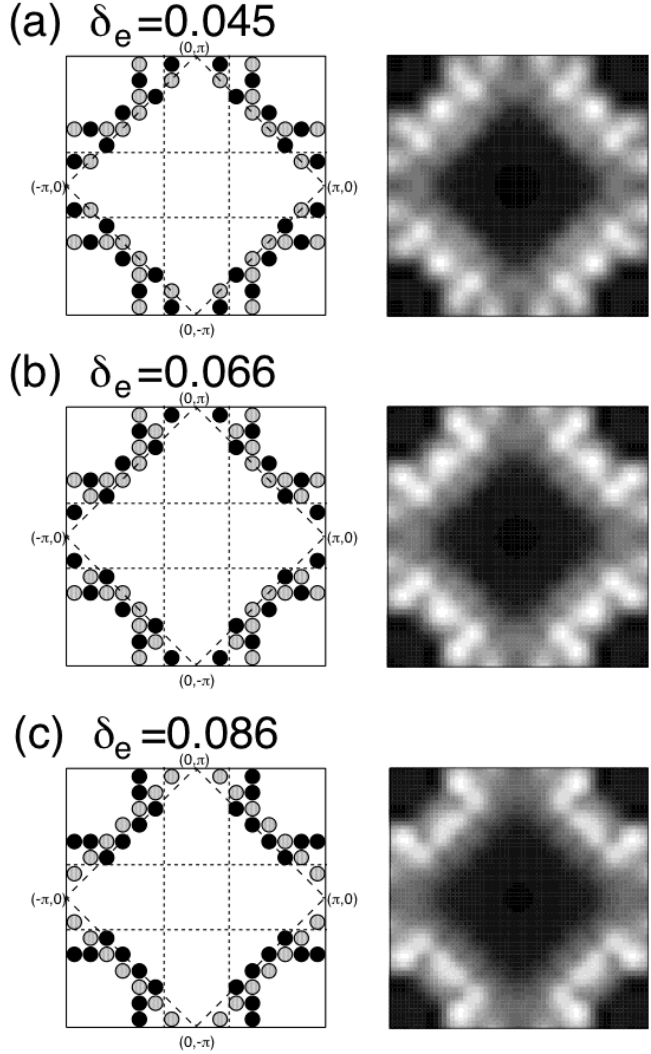}
\caption{Fermi surface evolutions for lightly-electron-doped states. (a) for $\delta_{\mathrm{e}}=0.045$. (b) for $\delta_{\mathrm{e}}=0.066$. (b) for $\delta_{\mathrm{e}}=0.086$.}
\label{figure:5}
\end {figure}
\begin{figure}
\includegraphics[width=8.6cm]{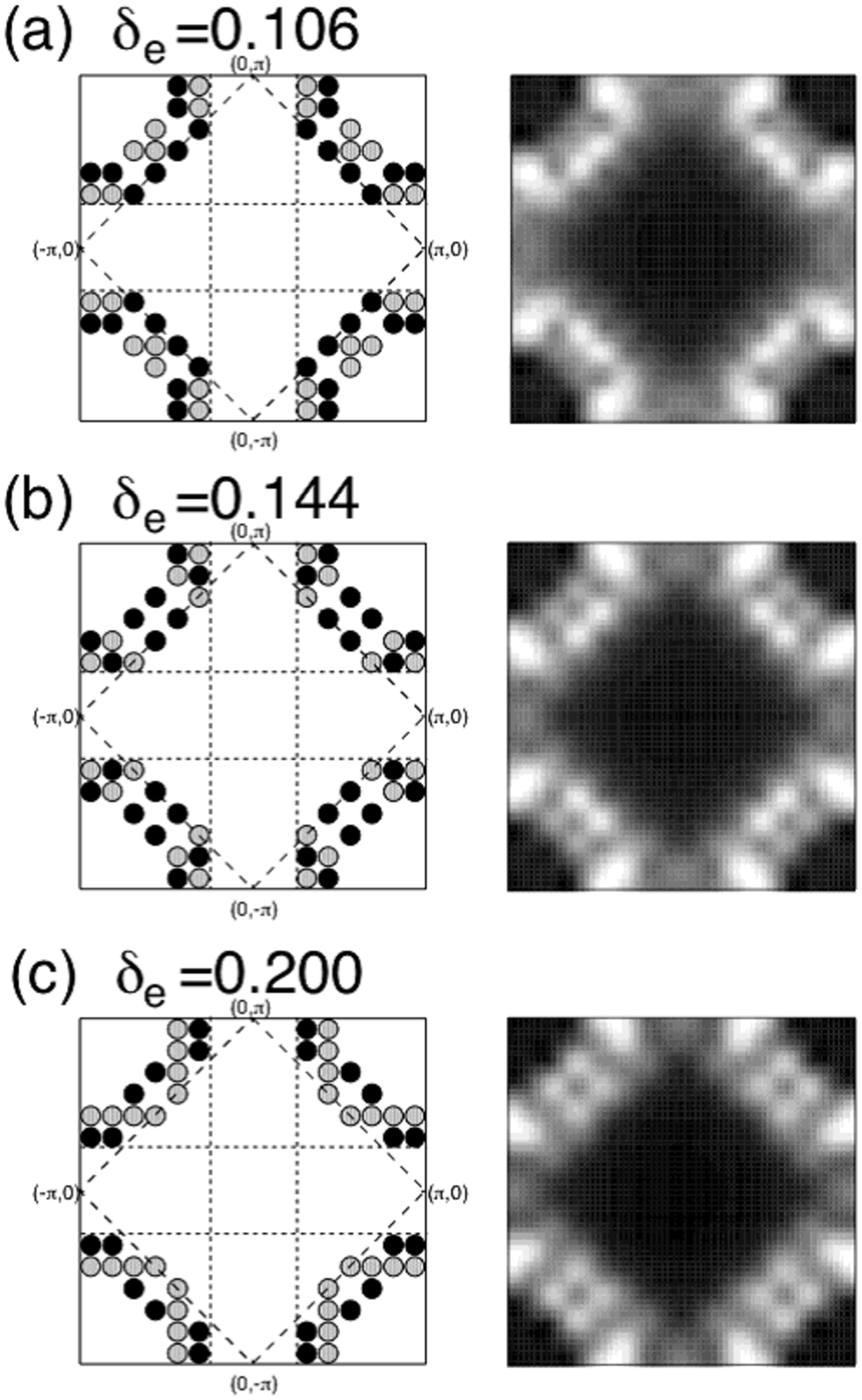}
\caption{Fermi surface evolutions for heavily-electron-doped states. (a) for $\delta_{\mathrm{e}}=0.106$. (b) for $\delta_{\mathrm{e}}=0.144$. (b) for $\delta_{\mathrm{e}}=0.200$.}
\label{figure:6}
\end{figure}
Secondly, we show our results for lightly-hole-doped states, 
in which the anti-ferromagnetic instability might exist. 
In this case we have no more eight empty spots but four empty spots 
centered on $(\pm\pi/2,\pm\pi/2)$ as shown in Fig.~\ref{figure:2}. These 
empty spots are called 'Fermi arc's, which is believed to appear when 
the strong anti-ferromagnetic fluctuation exists. In fact, these Fermi arcs 
have been observed in La$_{1.97}$Sr$_{0.03}$CuO$_4$ 
by ARPES and interpreted as the evidence of the 
anti-ferromagnetic long-range order.\cite{TYoshida2003} In our calculating 
results, the spectral weights in these Fermi arcs are not completely zero, and 
the anti-ferromagnetic long-range order does not exist. However, the existence 
of these Fermi arcs should be originated from 
the strong anti-ferromagnetic fluctuation, evolving 
toward the anti-ferromagnetic long-range order.
 
When the doped holes become less, electrons should 
localize due to strong coulomb repulsion among them in real materials. 
However, in our three-band Hubbard model electrons cannot localize even 
around near half-filling and the Fermi surface are remained, as 
shown in Fig.~\ref{figure:4}. If we could reproduce the Fermi surface 
evolution for the whole doping region, 
the one electron spectral weight at the 
Fermi level should be almost vanished near half-filling. 
This result means that our formulation on 
the basis of UFLEX is not efficient to describe the Mott-Hubbard 
metal-insulator transition. Unfortunately this is the 
side-effect of UFLEX, in which we overestimate the thermal fluctuations 
and recover the itinerancy of electrons. It would be our future issue to 
describe the localized electrons but with strong fluctuations appropriately. 

Our results for electron-doped states will 
be highlighted. The ARPES have already clarified the doping 
dependence of the Fermi surface in 
Nd$_{\rm 2-x}$Ce$_{\rm x}$CuO$_{4 \pm \delta}$.\cite{Armitage2002,Armitage2003,Shen2004} Except for completely vanished Fermi surface near half-filling, 
their observed evolution of Fermi surface with electron doping are consistent with our calculated result. As shown in Fig.~\ref{figure:5}, 
in lightly-electron-doped region the eight pockets emerge 
around the cross-sectional point of Fermi surface with FBZ boundaries. 
This reflects that the vertical charge fluctuation exists in the 
lightly-electron-doped region 
as well as near $1/8$-hole-doped region. We can guess that 
the vertical charge fluctuation gets to play an important role 
when the localized electron recovers its itinerancy 
as doped electron increases. 
When the doped-electron density density increases more, 
each two of eight pockets on the same 
Fermi surface section come close to each other and unite into one 
as shown in Fig.~\ref{figure:6}. Thus, 
in heavily-electron-doped region we have four pockets 
around the cross-sectional point of Fermi surface with $k_x=\pm k_y$. 
This means that the diagonal spin fluctuation, which is 
not completely anti-ferromagnetic, 
is strengthened instead of the vertical fluctuation. 
This diagonal spin fluctuation can cause pseudo-gap phenomena in electron-doped 
high-T$_{\rm c}$ as observed at low temperature.\cite{Onose2004} 
It contrasts with the case of the hole-doped region, and the difference 
should be caused by both a large $t_{pp}$. Such a $t_{pp}$ makes 
the big difference of the energy dispersions between the hole-doped 
region and the electron-doped region. If the electron were localized, 
the difference of the energy dispersions should 
appear as the different charge distributions. 
In our results the electron could not be localized, 
but the charge fluctuations appear in the different way between the hole-doped 
region and the electron-doped region instead. Hence, we can insist that 
the inhomogeneity of the strongly correlated system should be considered. 

\section{Spin and charge correlations}

In this section we show how the Fermi surface evolution 
with doping is related to the development 
of the spin and charge correlations. The momentum dependences of 
spin and charge correlation functions 
reflect the spin and charge spatial inhomogeneities, respectively. 

\begin{figure}
\includegraphics[width=8.6cm]{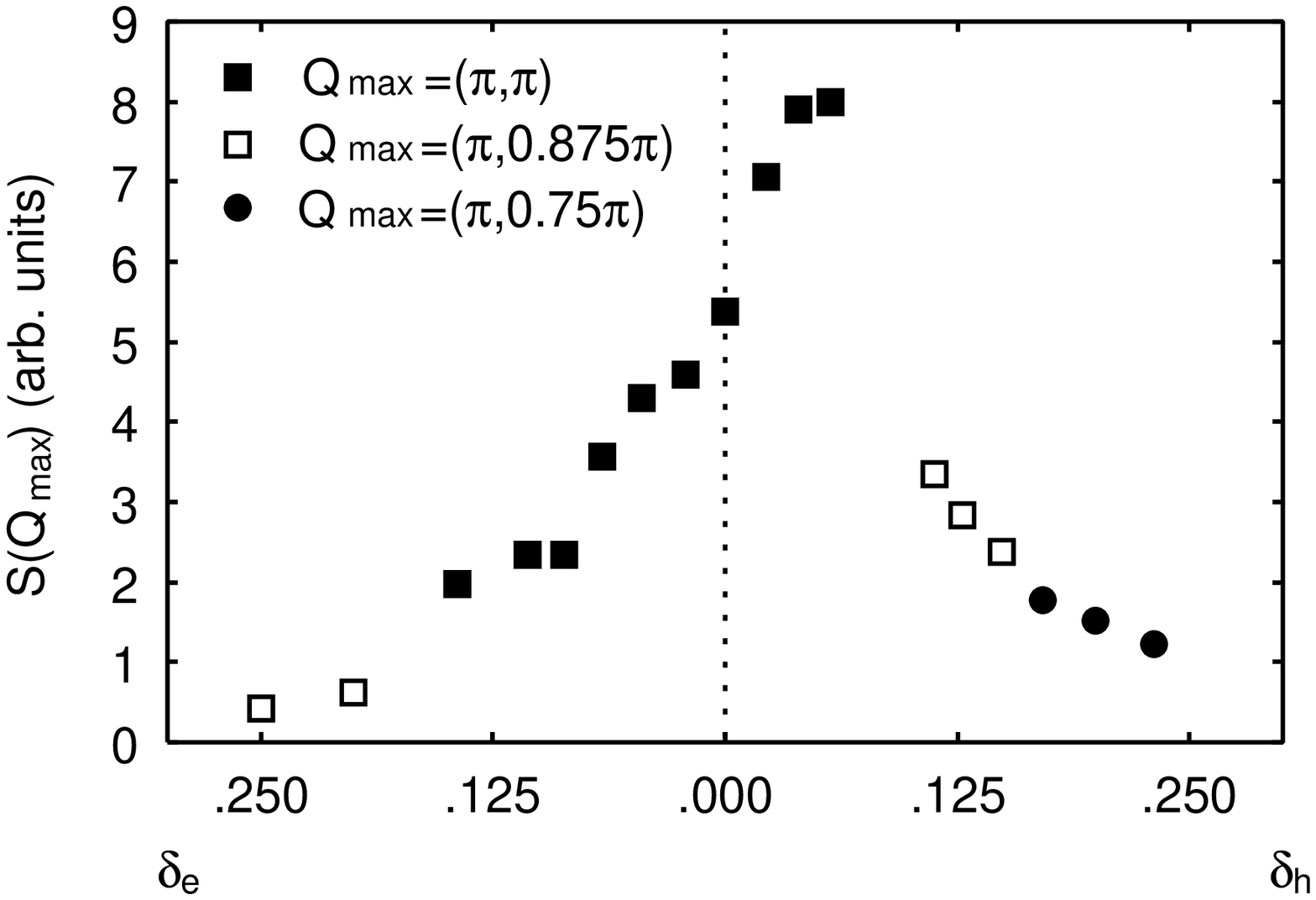}
\caption{The doping dependence of $S(\mathbf{q})$ at its maximum momentum 
$\mathbf{Q}_{\mathrm{max}}$.}
\label{figure:7}
\end{figure}
\begin{figure}
\includegraphics[width=8.6cm]{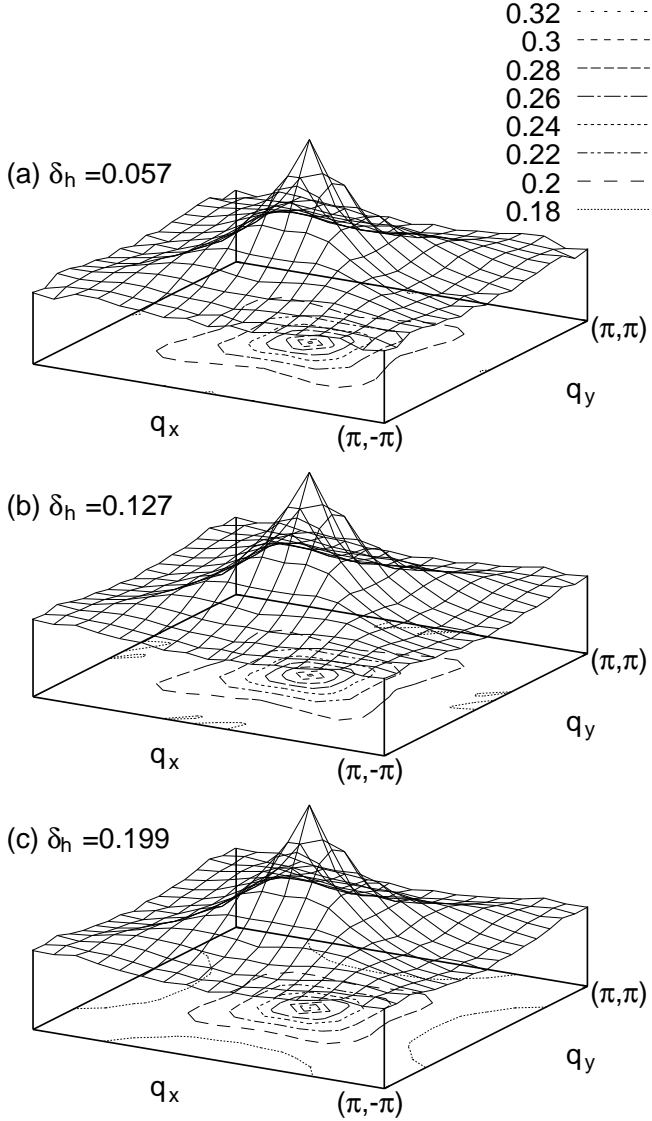}
\caption{The momentum dependences of $D(\mathbf{q})$ at hole-doped regions.}
\label{figure:8}
\end {figure}
\begin{figure}
\includegraphics[width=8.6cm]{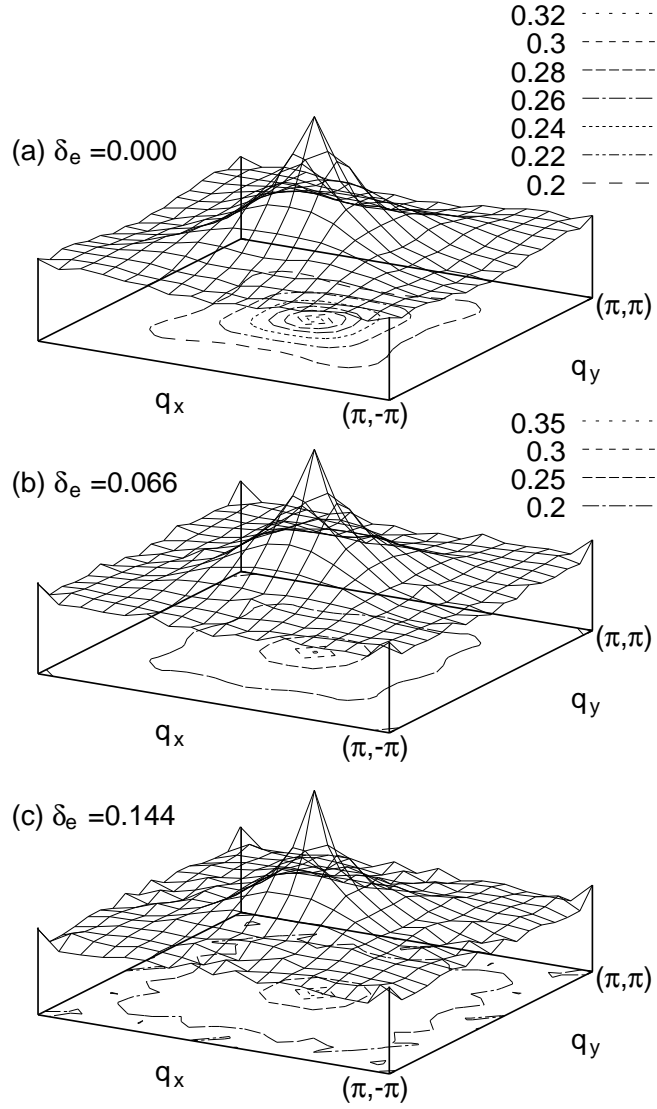}
\caption{The momentum dependences of $D(\mathbf{q})$ at electron-doped regions.}
\label{figure:9}
\end {figure}
At first we sum up our obtained results 
about the static spin correlations as both the incommensurabilities and 
the intensities of the peaks in the calculated elastic neutron scattering. 
They are shown in Fig.~\ref{figure:7}, where
\begin{equation}
S(\mathbf{q}) \equiv
\sum_{\mathbf{K}}
\chi_{-\mathbf{K}}^{+-}
(\mathbf{q},0)
\left[
\delta_{\mathbf{K}}
-U\chi_{\mathbf{K}}^{+-}(\mathbf{q}-\mathbf{K},0)
\right]^{-1}.
\end{equation}
In the hole-doped region the incommensurabilities $\delta_{\rm inc}$, 
defined by 
\begin{equation}
\delta_{\mathrm{inc}} 
\equiv \mathrm{Max}\{[\mathbf{Q}_{\mathrm{max}}-
(\pi,\pi)]_x,[\mathbf{Q}_{\mathrm{max}}-(\pi,\pi)]_y\},
\end{equation}
are almost proportional to $\delta_{\rm h}$. This has been already observed by  
the neutron scattering experiment on La$_{\mathrm{2-x-y}}$Ba$_{\mathrm{x}}$Sr$_{\mathrm{y}}$CuO$_4$.\cite{Yamada1998,Fujita2002PRB,Fujita2002PRL} The ratio of 
$\delta_{\rm inc}$ to $\delta_{\rm h}$ in our calculating results are almost 
half of the one in these experiments. However, we believe that this deviation 
is originated from the difference between the real hole number 
$\delta_{\rm h}$ introduced in CuO$_2$ plane, which is correspond to the 
one in our results, and the instoicheiometric number $\delta$ used in 
the analysis in Ref.~\citen{Yamada1998}.

On the other hand, in the electron-doped region 
we cannot recognize such a relationship between $\delta_{\mathrm{inc}}$ and 
$\delta_{\mathrm{e}}$. It suggests that our spin correlation functions do not 
simply reflect the Fermi surface nesting but rather the existence of 
inhomogeneities as discussed in Section~\ref{Fermi}.  

In order to clear our insistence, we calculate the static 
charge correlation functions at some regions. The momentum dependences 
of these correlation functions are shown in 
Figs.~\ref{figure:8} and \ref{figure:9}, where
\begin{equation}
D(\mathbf{q}) \equiv 
\frac{T}{N}\,\sum_{\mathbf{k} \mathbf{K}^\prime\, n}
G_{d\,\mathbf{K}^\prime}^{\,\sigma}(\mathbf{q}-\mathbf{k}, 
-\mathrm{i}\epsilon_n)
\,G_{d\,-\mathbf{K}^\prime}^{-\sigma}(\mathbf{k},\mathrm{i}\epsilon_n).
\end{equation}
Focusing on Fig.~\ref{figure:8}(b) and Fig.~\ref{figure:9}(b), 
we can recognize that $D(\mathbf{q})$ are slightly enhanced around the 
symmetric lines, $q_x=0$ and $q_y=0$. These momentum dependences of 
$D(\mathbf{q})$ suggest that the two collective modes 
with the momenta $(\pm 1,0)$ and $(0,\pm 1)$ exist. These collective modes 
can be translated as the instabilities toward 
vertical charge orderings along $x$-axis and $y$-axis, respectively. 
In fact, when $\delta_{\mathrm{h}}=0.127$ and 
$\delta_{\mathrm{e}}=0.066$, the Fermi surfaces have eight pockets around 
$(\pm\pi,\pm\pi/4)$ and $(\pm\pi/4,\pm\pi)$ as shown in Fig.~\ref{figure:1}(b) 
and Fig.~\ref{figure:4}(b), respectively. As discussed in \ref{Fermi}, 
these anomalous Fermi surfaces are caused from the charge inhomogeneities, 
which appear as shown in Fig.~\ref{figure:8}(b) and Fig.~\ref{figure:9}(b). 

%
%
\section{Conclusion}
 
In this paper, we have obtained fully self-consistent solutions 
for the 2D three-band Hubbard model on the basis of UFLEX, in 
which the inhomogeneous distribution of d-electrons are allowed. 
Both in hole-under-doped and electron-doped regions the one-particle 
spectral weights behave anomalously. Furthermore, 
the spin and charge correlation functions 
reflect the existence of super-lattice structures around 1/8-filling. 
Our numerical solutions show that the metallic state with spatially 
inhomogeneous spin or charge fast fluctuation can exist at high-temperature. 
The electronic state in one of our microscopically derived 
solutions corresponds to the dynamical stripe state mentioned in some 
pioneering works.\cite{Zaanen1996,Zaanen2000,Zaanen2003,Hasselmann2002} 
In our 2D model, the electrons cannot have 
the long-range order at finite temperature because of the strong 
fluctuation characteristic of low-dimensionality. However, 
in three-dimensional real materials, the spatially inhomogeneous fluctuation 
may tend to have the long-range order and form the striped state, 
observed in the neutron scattering experiments. This long-range order 
formation in three-dimensional real materials might not be reproduced 
by our simple model in this paper. The existences of the 
phonon or the long-range Coulomb potential might have an 
important role for these long-range order formation. 
The theoretical researches concerning these complicated factors are 
to be expected as the future problem.

\section*{Acknowledgments}
The authors are grateful to Professor K. Yamaji, Professor Y. Aiura, Professor H. Eisaki, Dr. I. Nagai, Dr. M. Miyazaki, and Dr. S. Koike for their invaluable comments. The computation in this work has been done using the facilities of the Supercomputer Center, Institute for Solid State Physics, University of Tokyo. The early development of the code for this computation has been achieved by using the IBM RS/6000--SP at TACC and VT-Alpha servers at NeRI in AIST.

\end{document}